# The Parameterized Complexity Analysis of Partition Sort for Negative Binomial Distribution Inputs

Niraj Kumar Singh[1], Mita Pal[2] and Soubhik Chakraborty[3]*

[1]Department of Computer Science & Engineering, B.I.T. Mesra, Ranchi-835215, India

[2,3]Department of Applied Mathematics, B.I.T. Mesra, Ranchi-835215, India

*email address of the corresponding author: soubhikc@yahoo.co.in
(S. Chakraborty)

**Abstract**

The present paper makes a study on Partition sort algorithm for negative binomial inputs. Comparing the results with those for binomial inputs in our previous work, we find that this algorithm is sensitive to parameters of both distributions. But the main effects as well as the interaction effects involving these parameters and the input size are more significant for negative binomial case.

## 1. Introduction

Partition Sort was introduced in [1] which indicates of a higher average case robustness compared to that of the popular quick sort algorithm [2]. This robustness was further reconfirmed in [3] where it was subjected to an unconventional distribution (Cauchy) inputs apart from the parameterized complexity analysis over binomial inputs. Here in this paper we study this algorithm for Negative Binomial, NB(k, p), inputs. Our first study suggests an empirical $O(n\log n)$ complexity for this distribution data. Next, as a parameterized complexity analysis, for different p, the probability of success, the average run times are observed and found to be a quadratic function of p, $Y_{avg}(n, k, p) = O_{emp}(p^2)$ for fixed n and k values. Further when k (the desired number of successes) is varied the average run times are found to be a quadratic function of k, $Y_{avg}(n, k, p) = O_{emp}(k^3)$ for fixed n and k values. Lastly, we focus on parameterized complexity, using factorial experiments, when the n observations to be sorted come from Negative Binomial population NB (k, p). To investigate the individual effect of number of sorting elements (n), negative binomial parameters and also their joint effects, a 3-cube factorial experiment is conducted with three levels of each of the factors n, k and p. Comparing the result of factorial experiment of binomial distribution inputs (see [3]) and the result of factorial experiment of negative distribution inputs, it is observed that negative binomial distribution is more sensitive than binomial distribution for all main effect and interactive effect also.

## 2. The Algorithm: Partition Sort

Introduced by Singh and Chakraborty [1], Partition sort is divide and conquer based a robust and efficient comparison sort algorithm. The key sub routine 'partition' when applied on input A[1…….n] divides this list into two halves of sizes floor (n/2) and ceiling (n/2) respectively. The property of the elements in these halves is such that the value of each element in first half is less than the value of every element in the second half. The recursive call to Partition-sort routine finally yields a sorted sequence of data as desired. The worst case performance of Partition Sort is found to be $O(n\log_2^2 n)$, whereas the best case count is $\Omega(n\log_2 n)$. The average case performance as estimated through the statistical bound is empirical $O(n\log_2 n)$ which is obtained by working directly on time. The reason for going with statistical bound may be found in [1] and [4].

## 3. Statistical Analysis

Negative Binomial (NB) distribution is obtained by performing independent Bernoullian trials ( a Bernoullian trial is one that results in one of two possible outcomes which we call 'success' and 'failure') till the desired number of successes say 'k' are obtained, with p as the constant probability of successes in a trial. The number of trials required is the NB variate with parameters k and p. This section includes empirical results performed over Partition Sort algorithm for negative binomial inputs to obtain the simplest model fitted to (time) complexity data that is neither an underfit nor an overfit [4]. The average case analysis is performed using statistical bound estimate (or empirical O). For definitions of statistical bound and 'empirical-O' refer to [1], [4]. Average case analysis was done by directly working on program run time to estimate the weight based statistical bound over a finite range by running computer experiments [5] [6].

The entry 'T' in the tables 1-3 denotes the mean time where each mean time (in sec.) data is averaged over 100 trial readings.

**Remark:** All the experiments have been carried out using PENTIUM 1600 MHz processor and 512 MB RAM.

### 3.1 Average Case Analysis Using Statistical Bound Estimate or Empirical O

The negative binomial distribution inputs are taken with parameters k and p, where k=1000 and p=0.5 are fixed. The empirical result is shown in fig (1). Experimental result as shown in fig (1) is suggesting a step function that is trying to get close to O(nlogn) complexity. So we can safely conclude that
$$Y_{avg}(n) = O_{emp}(nlogn).$$
The subscript "emp" implies an empirical and hence subjective bound-estimate [4].

Table 1: mean time (in sec.) for negative binomial distribution inputs, p=0.5, k=1000

| n | 10000 | 20000 | 30000 | 40000 | 50000 | 60000 | 70000 | 80000 | 90000 | 100000 |
|---|---|---|---|---|---|---|---|---|---|---|
| T | 0.02022 | 0.04282 | 0.06946 | 0.09474 | 0.11876 | 0.14932 | 0.1753 | 0.20566 | 0.23446 | 0.26462 |

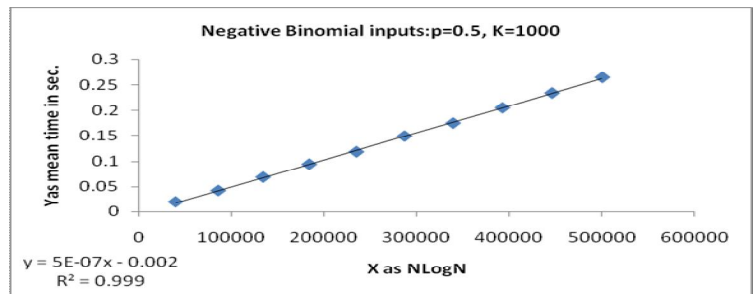

Fig 1: Regression model suggesting empirical O(nlogn) complexity

### 3.2 Parameterized Complexity Analysis

The Study of Parameterized complexity is an essential activity of any statistical analysis for accessing the true potential of an algorithm's performance. Our previous related work [3] suggests that for Partition Sort, the parameters of the input distribution should also be taken into account for explaining its complexity, and not just the parameter characterizing the size of the input. The study in this section is accordingly devoted to parameterized complexity analysis whereby the sorting elements of Partition Sort come independently from a Negative Binomial (k, p) distribution. Here our interest lies in investigating the response behavior (which is CPU time in our case) as a function of input distribution parameter(s). The first systematic work on parameterized complexity was done by Downey & Fellows [7]. Other significant work on this topic may be found in [8] and [9].

CASE (A): Parameterized Complexity Analysis when n and k are fixed while p is varying

The first analysis is done for fixed n and k values, while the p value is varied in the range [0.1 to 0.9]. The experimental result is put into Table 2 and the corresponding plots are given in figure 2.

Table 2: mean time (in sec.) for negative binomial distribution inputs, N=50000, k=1000

| p | 0.1 | 0.2 | 0.3 | 0.4 | 0.5 | 0.6 | 0.7 | 0.8 | 0.9 |
|---|---|---|---|---|---|---|---|---|---|
| T | 0.14674 | 0.13838 | 0.13438 | 0.12594 | 0.12066 | 0.11776 | 0.11028 | 0.10432 | 0.09524 |

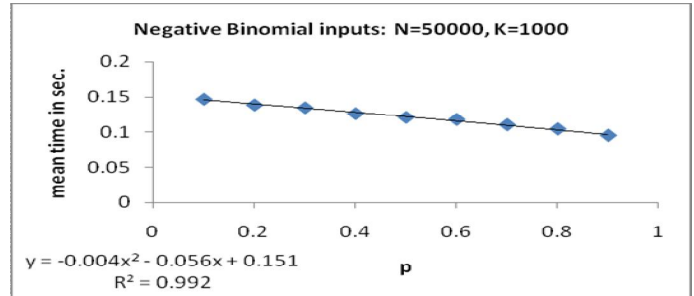

Fig 2: second degree polynomial fit

The experimental result suggests an average function $Y_{avg}(n, k, p) = O_{emp}(p^2)$ for fixed n and k values.

CASE (B): Parameterized Complexity Analysis when n and p are fixed while k is varying

The next analysis is done for fixed n and p values, while the k value is varying in the range [100 to 5000]. The experimental result is put into Table 3 and the corresponding plots are given in figure 3.

Table 3 mean time (in sec.) for negative binomial distribution inputs, N=50000, p=0.5

| K | 100 | 500 | 1000 | 2000 | 3000 | 4000 | 5000 |
|---|---|---|---|---|---|---|---|
| T | 0.1001 | 0.11398 | 0.11876 | 0.12594 | 0.12842 | 0.13098 | 0.13344 |

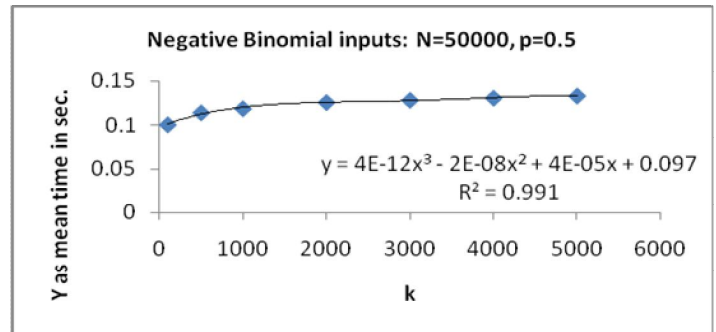

Fig 3: Third degree polynomial fit for varying k values

The experimental result suggests an average function $Y_{avg}(n, k, p) = O_{emp}(k^3)$ for fixed n and k values.

## Partition Sort subjected to Factorial Experiments

Factorials experiments are experiments that investigate the effects of two or more factors or input parameters on the output response of a process. Factorial experiment design, or simply factorial design, is a systematic method for formulating the steps needed to successfully implement a factorial experiment. Estimating the effects of various factors on the output of a process with a minimal number of observations is crucial to being able to optimize the output of the process.

We further performed the parameterized complexity analysis by conducting a 3-cube factorial experiment with three levels of each of the three factors n, m and p. All the three factors are found to be significant both individually and interactively. Table 4 contains the data for $3^3$ factorial experiments for Partition Sort when input distribution is negative binomial.

Table 4: Partition sort times in second Binomial (k, p) distribution input for various n (10000, 30000, 50000), k (1000, 3000, 5000) and p (0.2, 0.5, 0.8).

| P=0.2 | | | |
|---|---|---|---|
| N | k=1000 | k=3000 | k=5000 |
| 10000 | 0.01966 | 0.02324 | 0.02376 |
| 30000 | 0.0778 | 0.07878 | 0.08214 |
| 50000 | 0.13844 | 0.1415 | 0.14746 |
| P=0.5 | | | |
| N | k=1000 | k=3000 | k=5000 |
| 10000 | 0.02022 | 0.02164 | 0.02094 |
| 30000 | 0.06946 | 0.0723 | 0.075 |
| 50000 | 0.11876 | 0.12842 | 0.13344 |
| P=0.8 | | | |
| N | k=1000 | k=3000 | k=5000 |
| 10000 | 0.0172 | 0.01932 | 0.01832 |
| 30000 | 0.05954 | 0.06334 | 0.06676 |
| 50000 | 0.1031 | 0.11278 | 0.11784 |

**Result for $3^3$ factorial experiment**

Factorial experiment is performed using MINITAB statistical package version 15. The analysis data obtained is put in the following result.

**Multilevel Factorial Design**

```
Factors:       3    Replicates:      3
Base runs:    27    Total runs:     81
Base blocks:   1    Total blocks:    1

Number of levels: 3, 3, 3
```

**General Linear Model: y versus n, p, k**

```
Factor  Type   Levels  Values
n       fixed       3  1, 2, 3
p       fixed       3  1, 2, 3
k       fixed       3  1, 2, 3
```

Analysis of Variance for y, using Adjusted SS for Tests

```
Source  DF     Seq SS     Adj SS     Adj MS            F      P
n        2  0.1528421  0.1528421  0.0764211  1.77367E+08  0.000
p        2  0.0039854  0.0039854  0.0019927   4624925.02  0.000
k        2  0.0006386  0.0006386  0.0003193    741021.05  0.000
n*p      4  0.0016832  0.0016832  0.0004208    976642.57  0.000
n*k      4  0.0002823  0.0002823  0.0000706    163825.64  0.000
p*k      4  0.0000188  0.0000188  0.0000047     10901.72  0.000
n*p*k    8  0.0000517  0.0000517  0.0000065     14988.37  0.000
Error   54  0.0000000  0.0000000  0.0000000
Total   80  0.1595021
```

```
S = 0.0000207573   R-Sq = 100.00%   R-Sq(adj) = 100.00%
```

Experimental results reveal that Partition sort is highly affected by the main effects n, p and k. It is interesting to note that all interactions are found significant in Partition Sort. Moreover, it is observed that negative binomial distribution is more sensitive than binomial distribution [3] for all main effects and interaction effects as well. Interestingly the main effect p in negative binomial is more sensitive than p in binomial inputs.

## 4. Conclusion and suggestions for future work

Our experimental result and its subsequent analysis reveals that the Partition Sort exhibits robustness in the average case for negative binomial inputs and hence can serve as a better alternative than the popular quick sort algorithm. Apart from the robustness issue we also found it to be sensitive to input distribution parameters and hence a potential candidate to the study of parameterized complexity analysis. For n independent Negative Binomial (k, p) inputs, all the three factors are significant both independently and interactively. All the two factor interactions n*k, n*p and k*p and even the three factor n*k*p is significant. Moreover, it is observed that negative binomial distribution is more sensitive than binomial distribution [3] for all main effects and interaction effects also. Our finding regarding the measure of parametric influence is an experimental approach. Although this measure can be accomplished through some suitable theoretical analysis, we still are dependent on statistical tools so as to confirm the significance of their interactions.